\documentclass[aps,showpacs,preprintnumbers,nofootinbibt, twocolumn]{revtex4}
\usepackage{graphicx}
\usepackage{amsmath}
\usepackage{amssymb}
\usepackage{amsfonts}
\usepackage{bm}
\usepackage{color}

\def\be{\begin{equation}}
\def\ee{\end{equation}}
\def\bea{\begin{eqnarray}}
\def\eea{\end{eqnarray}}

\begin{document}

\title{Geodesic deviation, Raychaudhuri equation, and tidal forces in modified gravity with an arbitrary curvature-matter coupling}

\author{Tiberiu Harko}
\email{harko@hkucc.hku.hk} \affiliation{Department of Mathematics, University College London, Gower Street, London WC1E 6BT, United Kingdom}
\author{Francisco S.~N.~Lobo}\email{flobo@cii.fc.ul.pt}
\affiliation{Centro de Astronomia e Astrof\'{\i}sica da Universidade de Lisboa, Campo Grande, Ed. C8 1749-016 Lisboa, Portugal}

%% REVTEX4

\begin{abstract}

The geodesic deviation equation, describing the relative accelerations of
nearby particles, and the Raychaudhury equation, giving the evolution of the
kinematical quantities associated with deformations (expansion, shear
and rotation) are considered in the framework of modified theories of gravity with an arbitrary curvature-matter coupling, by taking into account the effects of the extra force. As a physical
application of the geodesic deviation equation the modifications of the
tidal forces due to the supplementary curvature-matter coupling are
obtained in the weak field approximation. The tidal motion of test particles is directly influenced not
only by the gradient of the extra force, which is basically determined by
the gradient of the Ricci scalar, but also by an explicit coupling between
the velocity and the Riemann curvature tensor. As a specific example, the expression of the Roche limit (the orbital distance at which a satellite will begin to be tidally torn apart by the body it is orbiting) is also obtained for this class of models.

\end{abstract}

\date{\today}


\maketitle

%%%%%%%%%%%%%%%%%%%%%%%%%%%%%%%%%%%%%%
\section{Introduction}
%%%%%%%%%%%%%%%%%%%%%%%%%%%%%%%%%%%%%%

The strong observational evidence that recently our universe has entered an
accelerated expansion phase \cite{Ri98}, and the astonishing result that
around $95$--$96\%$ of the content of the Universe is in the form of dark
energy and dark matter, respectively, with only about $4$--$5\%$ being
represented by baryonic matter \cite{PeRa03}, has shown the limitations of standard general relativity. Hence, despite the remarkable success of
general relativity at the solar system scale, on a galactic and cosmological
scale gravitational theories face two fundamental problems: the dark matter
problem, and the dark energy problem, respectively. Although in recent years
many different approaches have been proposed to explain the observational
results of cosmology, a satisfactory model has yet to be obtained.

However, a promising way to explain the observational data is to assume that at large scales the Einstein gravity model of general relativity breaks down,
and a more general action describes the gravitational field. Theoretical
models in which the standard Einstein-Hilbert action is replaced by an
arbitrary function of the Ricci scalar $R$, first proposed in \cite{Bu70},
have recently been extensively investigated. Cosmic acceleration can be
explained by $f(R)$ gravity \cite{Carroll:2003wy}, and viable  cosmological models can be found \cite{viablemodels}. For a review of $f(R)$ generalized gravity models see \cite{SoFa08}. The possibility that the galactic dynamic of massive test particles can be understood, without the need for dark matter, was also extensively considered in the framework of $f(R)$ gravity \cite{darkmat}.
In the context of the Solar System regime, local tests in the weak field approximation are also to be retained by assuming the chameleon mechanism \cite{chameleon}, where the mass of the chameleon scalar field depends on the local background matter density, i.e., in regions of high matter density the chameleon scalar field is massive. However, it has recently been shown that in the hybrid metric-Palatini gravitational theory, which consists of the superposition of the metric Einstein-Hilbert Lagrangian with an $f(R)$ term constructed \`a la Palatini \cite{Harko:2011nh}, that even if the scalar field is very light, the theory passes the Solar System observational constraints. Therefore the model predicts the existence of a long-range scalar field, modifying the cosmological and galactic dynamics.

However, most of the generalizations of standard general relativity
concentrated only on the geometric part of the action, and assumed that the
matter part is unchanged, that is, in the total Lagrangian the matter term
was considered as a simple additive term. This point of view severely limits
the possibilities of the matter-geometry interaction, restricting
the degrees of freedom of the gravitational theories. From a physical point
of view, generalized gravitational models, involving curvature-matter
interactions, cannot be ruled out \textit{a priori} \cite{od}.
In this context, a generalization of $f(R)$ gravity was introduced in \cite{Bertolami:2007gv}, and extended in \cite{Ha08}, by including in the theory an explicit coupling of an arbitrary function of the curvature scalar, $R$, with the matter Lagrangian density. As a result of the coupling, the motion of the massive particles is non-geodesic, and an extra force, orthogonal to the four-velocity, arises. This class of models of modified gravity can be denoted as generalized gravity models with a linear curvature-matter coupling, and they have been extensively studied recently \cite{lit}. For a review of modified gravity models with curvature-matter coupling see \cite{Lobo}.

Similar couplings between gravitation and matter have also been considered,
as possible explanations for the accelerated expansion of the universe and
of the dark energy, in \cite{Od}.
An interesting extension of standard general relativity has also been been proposed, namely, the $f(R, T)$ modified theories of gravity, where the gravitational Lagrangian is given by an arbitrary function of the Ricci scalar $R$ and of the trace of the stress-energy tensor $T$ \cite{Harko:2011kv}. It is interesting to note that the dependence from $T$ may be induced by exotic imperfect fluids or quantum effects (conformal anomaly). This extended theory of gravity may be considered as a relativistically covariant model of interacting dark energy. It was further argued that the new matter and time dependent terms in the gravitational field equations play the role of an effective cosmological constant.

The above models were further generalized in a recent proposal where the gravitational action is given by an arbitrary function of the Ricci scalar $R$ and of the Lagrangian density of the matter $L_m$ \cite{HaLo08}, and with a generalized scalar field and kinetic term dependences \cite{Harko:2012hm}.
Thus, $f(R,L_m)$ represents the natural generalization of the models with linear matter coupling, as well as the most general extension of the standard Hilbert action for the gravitational field, $S=\int \left[ R+L_{m}\right] \sqrt{-g} d^{4}x$. In this class of models the energy-momentum tensor of the matter is generally not conserved, and the motion is non-geodesic. In the particular case in which the Lagrange function of the matter is a function of the energy density of the matter only, the equations of motion of test particles can be obtained, by using a variational principle.

It is the purpose of the present paper to investigate some other interesting properties of the motion of test particles in gravity models with an arbitrary
curvature-matter coupling. The weak field limit of the model is carefully analyzed, and it is shown that in first order in both Ricci scalar and matter energy density one obtains the Newtonian Poisson equation in the presence of an effective cosmological constant. The equation of
geodesic deviation and the Raychaudhury equation are also formulated by explicitly including in the equations the effects of the curvature-matter coupling and of the extra force. Some of the physical implications of the geodesic deviation equation, namely, the problem of the tidal forces in this class of models is also considered, and the generalization of the Roche limit is obtained.

The present paper is organized as follows. The field equations and the
equations of motion of the model are derived in Section~\ref{sect2}.  The geodesic deviation equation (the Jacobi equation) of the test particles, as well as the Raychaudhury equation of the model, are obtained in Section~\ref{sect3}. Some physical applications of the geodesic deviation equation are discussed in Section~\ref{sect4}. We discuss and conclude our results in Section~\ref{sect5}. In the present paper we consider a system of units with $8\pi G=c=1$, and we follow the Landau-Lifshitz \cite{LaLi} conventions for the metric signature and the definition of the curvature tensor.

%%%%%%%%%%%%%%%%%%%%%%%%%%%%%%%%%%%%%%%%%%%%%%%%%%%%%%%%%%%%%%%%%%%%%%%
\section{Field and motion equations with an arbitrary curvature-matter coupling}\label{sect2}
%%%%%%%%%%%%%%%%%%%%%%%%%%%%%%%%%%%%%%%%%%%%%%%%%%%%%%%%%%%%%%%%%%%%%%%

The most general action for a $f\left(R,L_m\right)$ type modified theory of
gravity involving an arbitrary coupling between matter and curvature is given
by \cite{Ha08}
\begin{equation}
S=\int f\left(R,L_m\right) \sqrt{-g}\;d^{4}x~,
\end{equation}
where $f\left(R,L_m\right)$ is an arbitrary function of the Ricci scalar $%
R$, and of the Lagrangian density corresponding to matter, $L_{m}$. The only requirement for the function $f\left(R,L_m\right)$ is to be an analytical function of $R$ and $L_{m}$, respectively, that is, it must possess a Taylor series expansion about any point. The matter energy-momentum tensor $T_{\mu \nu}$ is defined as
\begin{equation}
T_{\mu \nu }=-\frac{2}{\sqrt{-g}}\frac{\delta \left( \sqrt{-g}L_{m}\right) }{%
\delta g^{\mu \nu }}.
\end{equation}

By assuming that the Lagrangian density $L_{m}$ of the matter depends only
on the metric tensor components, and not on its derivatives, we obtain $%
T_{\mu \nu }=L_{m}g_{\mu \nu }-2\partial L_{m}/\partial g^{\mu \nu }$.

Varying the action with respect to the metric tensor $g_{\mu \nu }$ we
obtain the field equations of the model as
%\begin{widetext}
\bea
f_{R}\left( R,L_{m}\right) R_{\mu \nu }+ \hat{P}_{\mu \nu } f_{R}\left( R,L_{m}\right) -\frac{1}{2}\big[
f\left( R,L_{m}\right)
    \nonumber  \\
 -f_{L_{m}}\left( R,L_{m}\right)L_{m}\big] g_{\mu \nu }=
\frac{1}{2}%
f_{L_{m}}\left( R,L_{m}\right) T_{\mu \nu } \label{feq}  ,
\eea
%\end{widetext}
where we have denoted  $f_{R}\left( R,L_{m}\right) =\partial f\left(
R,L_{m}\right) /\partial R$ and $f_{L_m}\left( R,L_{m}\right) =\partial f\left(
R,L_{m}\right) /\partial L_{m}$, respectively, and we have introduced the
operator $\hat{P}_{\mu \nu }$, defined as
\begin{equation}
\hat{P}_{\mu \nu }=g_{\mu \nu }\square -\nabla _{\mu }\nabla _{\nu },
\end{equation}
with $\square =\nabla _{\mu }\nabla ^{\mu }$. The operator $\hat{P}_{\mu \nu }$ has the property $\hat{P}_{\mu }^{ \mu }=3\square $.

By contracting the field equations Eq.~(\ref{feq}), we obtain the scalar
equation
\begin{equation}
3\square f_R +f_R R-2f =\left(\frac{1}{2}T-2L_m\right)f_{L_m} ,  \label{contr}
\end{equation}
where $T=T_{\mu }^{\mu }$ is the trace of the matter energy-momentum tensor.
By eliminating the term $\square f_R\left( R,L_{m}\right) $ between
Eq.~(\ref{feq}) and Eq.~(\ref{contr}), we can reformulate the field
equations as
%\begin{widetext}
\begin{eqnarray}
R_{\mu \nu }=\Lambda \left( R,L_{m}\right) g_{\mu \nu }+\frac{1}{f_R\left(
R,L_{m}\right) }\nabla _{\mu }\nabla _{\nu }f_R\left( R,L_{m}\right)
   \nonumber  \\
+\Phi \left( R,L_{m}\right) \left( T_{\mu \nu }-\frac{1}{3}Tg_{\mu \nu }\right) ,
\label{eqrmunu}
\end{eqnarray}
%\end{widetext}
where we have denoted
\begin{equation}
\Lambda \left( R,L_{m}\right) =\frac{2f_R\left( R,L_{m}\right)R
-f\left( R,L_{m}\right) +f_{L_m}\left(R,
L_{m}\right)  L_{m}}{6f_R\left( R,L_{m}\right) },
\label{lambda}
\end{equation}
and
\begin{equation}
\Phi \left( R,L_{m}\right) =\frac{f_{L_m}\left(R, L_{m}\right)
}{f_R\left( R,L_{m}\right) },
\end{equation}
respectively.

By taking the covariant divergence of Eq.~(\ref{feq}),  we obtain for the divergence of the energy-momentum tensor $T_{\mu \nu}$ the following relationship
\begin{eqnarray}
\nabla ^{\mu }T_{\mu \nu }&=&\nabla ^{\mu }\ln \left[
f_{L_m}\left(R,L_m\right)\right] \left( L_{m}g_{\mu \nu
}-T_{\mu \nu }\right)
   \nonumber\\
&=& 2\nabla
^{\mu }\ln \left[ f_{L_m}\left(R,L_m\right) \right] \frac{\partial L_{m}}{%
\partial g^{\mu \nu }}.  \label{noncons}
\end{eqnarray}

Now, assuming that the matter Lagrangian is a function of the rest mass density $\rho $ of the matter only, from Eq.~(\ref{noncons}) we obtain explicitly the equation of motion of the test particles in the $f\left(R,L_m\right)$ gravity model as
\begin{equation}
\frac{D^{2}x^{\mu }}{ds^{2}}=U^{\nu }\nabla _{\nu }U^{\mu }=\frac{d^{2}x^{\mu }}{ds^{2}}+\Gamma _{\nu \lambda }^{\mu }U^{\nu }U^{\lambda
}=f^{\mu },
\label{eqgeod}
\end{equation}
where the wordline parameter $s$ is taken as the proper time, $U^{\mu
}=dx^{\mu }/ds$ is the four-velocity of the particle, $\Gamma
_{\sigma \beta }^{\nu }$ are the Christoffel symbols associated to the
metric, and the extra-force $f^{\mu}$ is defined as
\begin{equation}
f^{\mu }=-\nabla _{\nu }\ln \left[  f_{L_m}\left(R,L_m\right) \frac{%
dL_{m}\left( \rho \right) }{d\rho }\right] \left( U^{\mu }U^{\nu }-g^{\mu
\nu }\right) .
\label{extra}
\end{equation}

The extra-force $f^{\mu }$, generated by the curvature-matter coupling,  is perpendicular to the four-velocity, $f^{\mu}u_{\mu }=0$. Due to the presence of the extra-force $f^{\mu }$, the motion of the test particles in modified theories of gravity with an arbitrary coupling between matter and curvature is non-geodesic. From the relation $U_{\mu }\nabla _{\nu }U^{\mu }\equiv 0$ it follows that the force $f^{\mu }$ is always perpendicular to the velocity, so that $U_{\mu }f^{\mu }=0$.

The generalized equations of motion Eq.~(\ref{eqgeod}) can be derived from the Kahil-Bazanski Lagrangian
\cite{kahil},
\begin{equation}
L_{p}=g_{\mu \nu }U^{\mu }\frac{D\eta ^{\nu }}{ds}+f_{\mu }\eta ^{\mu
}=g_{\mu \nu }U^{\mu }\dot{\eta}^{\nu }+f_{\mu }\eta ^{\mu },
\end{equation}
where $\eta ^{\nu }$ is an arbitrary four-vector, which can be taken, for
example, as the deviation vector (see Section~\ref{sect4}), and we have
denoted, for simplicity,
\begin{equation}
\frac{D\eta ^{\nu }}{ds}=\frac{d\eta ^{\nu }}{ds}+\Gamma _{\sigma
\beta }^{\nu }\eta ^{\sigma }U^{\beta }=\dot{\eta}^{\nu }.
\end{equation}
 Then we obtain immediately $\partial L_{p}/\partial \dot{\eta}%
^{\sigma }=U_{\sigma }$, \ and $\partial L_{p}/\partial \eta ^{\sigma
}=\Gamma _{\sigma \beta }^{\nu }U_{\nu }U^{\beta }+f_{\sigma }$,
respectively. Finally, the Lagrange equations
\begin{equation}
\frac{d}{ds}\left( \frac{\partial
L_{p}}{\partial \dot{\eta}^{\nu }}\right) -\frac{\partial L_{p}}{\partial \eta
^{\nu }}=0,
\ee
provide the equations of motion Eq.~(\ref{eqgeod}).

%%%%%%%%%%%%%%%%%%%%%%%%%%%%%%%%%%%%%%%%%%%%%%%%%%%%%%%%%%%%%%%%%%%
\section{Weak field limit  of the field equations in $f\left(R,L_m\right)$ gravity}\label{sect3}
%%%%%%%%%%%%%%%%%%%%%%%%%%%%%%%%%%%%%%%%%%%%%%%%%%%%%%%%%%%%%%%%%%%

Generally the matter Lagrangian $L_{m}$ \ is a function of the matter energy
density $\rho $, the pressure $p$ as well as the other thermodynamic
quantities, such as the specific entropy $s$ or the baryon number $n$, so that $L_{m}=L_{m}\left( \rho ,p,s,n\right) $. In the simple (but physically the
most relevant) case in which the matter obeys a barotropic equation of
state, so that the pressure is a function of the energy density of the
matter only, $p=p\left( \rho \right) $, the matter Lagrangian becomes a
function of the energy density only, and hence $L_{m}=L_{m}\left( \rho
\right) $. Then, the matter Lagrangian is given by \cite
{Ha08}
\begin{equation}\label{lagr}
L_{m}\left( \rho \right) =\rho \left(1+\int_{0}^{p}\frac{dp}{\rho }\right)-p\left( \rho \right),
\end{equation}
while the energy-momentum tensor can be written as
\begin{equation}\label{tens1}
T^{\mu \nu }=\left[ \rho +p\left( \rho \right) +\rho \Pi \left( \rho \right) %
\right] U^{\mu }U^{\nu }-p\left( \rho \right) g^{\mu \nu },
\end{equation}
respectively, where
\begin{equation}
\Pi \left( \rho \right) =\int_{0}^{\rho }\frac{p}{\rho ^{2}}d\rho
=\int_{0}^{p}\frac{dp}{\rho }-\frac{p\left(\rho \right)}{\rho }.
\end{equation}

 The expression $\Pi(\rho) + p(\rho)/\rho$ represents the specific enthalpy of the fluid. From a physical point of view $\Pi \left(\rho \right)$ can be interpreted as the elastic (deformation) potential energy of the body, and therefore Eq.~(\ref{tens1}) corresponds to the energy-momentum tensor of a compressible elastic isotropic system.

Next, we consider the weak field limit of the gravitational field equations. First, we assume that $\rho \gg p$, and therefore we systematically neglect the pressure term. Then, from Eq.~(\ref{lagr}) it follows that $L_m=\rho $, so that the energy-momentum tensor is given by  $T_{\mu \nu }=\rho U_{\mu }U_{\nu }$.

Secondly, we consider non-relativistic macroscopic motion. Consequently we can neglect the spatial components in the four-velocity, and retain only the time component, so that $U^{\mu }=U_{\mu }\approx(1,0,0,0)$. Since in the weak field limit one can omit all time derivatives as well as the terms containing the products of the Christoffel symbols, we obtain for $R_{00}$ the expression \cite{LaLi}
\be
R_{00}=R_0^0=-\frac{\eta^{il}\partial ^{2}\phi }
{\partial x^{i}\partial x^{l}}=\Delta \phi .
\ee
Similarly, we have $\Gamma _{0,i0}=\Gamma ^0_{i0}=\partial \phi /\partial x^{i}$, $i=1,2,3$ and $R_{i}^{i}=\partial ^{2}\phi
/\left( \partial x^{i}\right) ^{2}$, $i=1,2,3$ (no summation upon the index $i$). Therefore we obtain $R\approx 2\Delta \phi $.

Now, multiplying Eq.~(\ref{eqrmunu}) with $U^{\mu }U^{\nu }$ gives the scalar
equation
%\begin{widetext}
\begin{eqnarray}
U^{\mu }U^{\nu }R_{\mu \nu }=\frac{1}{f_R\left(
R,L_{m}\right) }U^{\mu }U^{\nu }\nabla _{\mu }\nabla _{\nu }f_R\left(
R,L_{m}\right)
     \nonumber  \\
+\Lambda \left( R,L_{m}\right)+\Phi \left( R,L_{m}\right) \left( U^{\mu }U^{\nu }T_{\mu \nu
}-\frac{1}{3}T\right) .  \label{scal}
\end{eqnarray}
%\end{widetext}

In the weak field limit and for a static geometry,  Eq.~(\ref{scal}) immediately provides the following relationship
\begin{equation}
\Delta \phi =\Lambda \left( R,L_{m}\right) +\frac{2}{3}\Phi \left(
R,L_{m}\right) \rho .  \label{poi1}
\end{equation}

The left hand side of Eq.~(\ref{poi1}) is first order in $1/c^2$. Therefore, we will estimate the right hand side of Eq.~(\ref{poi1}) in the same order of approximation. By using a Taylor series expansion to first order of approximation we obtain first for the function $\Lambda \left(R,L_m\right)$ the approximate representation
\be
 \Lambda \left(R,L_m\right)=\alpha +\beta R +\gamma L_m,
 \ee
 where
 \be
 \alpha =-\frac{f(0,0)}{6 f_R(0,0)},
 \ee
 \be
 \beta =\frac{ f_{R}^3(0,0)+f(0,0) f_{RR}(0,0) f_{R}(0,0)}{6
   f_{R}^3(0,0)},
  \ee
  and
  \be
  \gamma =\frac{f_{RL_m}(0,0) f(0,0)}{6
   f_{R}^2(0,0)},
\ee
respectively.

For the function $(2/3)\Phi \left( R,L_{m}\right)\rho $ we
obtain
\be
\frac{2}{3}\Phi \left( R,L_{m}\right)\rho \approx \frac{2}{3}\delta \rho ,
\ee
where
\be
\delta =\frac{f_{L_m}(0,0)}{f_{R}(0,0)}.
\ee

With these approximations Eq.~(\ref{poi1}) becomes
\begin{equation}\label{poi3}
\Delta \phi \approx \alpha +\beta R+\left(\gamma +\frac{2}{3}\delta \right)\rho .
\end{equation}
By substituting into Eq.~(\ref{poi3}) the weak field approximation of $R$, $R\approx 2\Delta \phi $, gives the generalized Poisson equation in modified theories of gravity with an arbitrary matter-geometry coupling as
\begin{equation}
\Delta \phi \approx \frac{\gamma +2\delta /3}{1-2\beta }\rho +\Lambda _0,  \label{poi4}
\end{equation}
where the condition $ \beta\neq 1/2$ must hold for all $\rho $, and
where we have denoted $\Lambda _0=\alpha /\left(1-2\beta \right)$.

In order to obtain the correct limit of the Newtonian
Poisson equation, the function $f$ and its derivatives estimated at the point $(0,0)$  must satisfy the
condition $\left(\gamma +2\delta /3\right)/\left(1-2\beta \right) =1/2$. The constant $\Lambda _0$
plays the role of an effective cosmological constant, which is naturally generated in the present model. In most of the
astrophysical applications $\Lambda _0$  can be neglected. Therefore in the first approximation  for the potential of the gravitational field of a single
particle of mass $m$ we obtain $\phi \left( r\right) =-m/8\pi r$, which is
the expression of the standard Newtonian potential. As a result, in the equation of motion of the test particles one can take $\vec{a}=-\nabla \phi =\vec{a}_N$, where $\vec{a}_N$ is the Newtonian acceleration of the particle.

%%%%%%%%%%%%%%%%%%%%%%%%%%%%%%%%%%%%%%%%%%%%%%%%%%%%%%%%%%%%%%%%%%%%%%%%%%%
\section{Geodesic deviation and the Raychaudhury equation with an
arbitrary curvature-matter coupling}\label{sect4}
%%%%%%%%%%%%%%%%%%%%%%%%%%%%%%%%%%%%%%%%%%%%%%%%%%%%%%%%%%%%%%%%%%%%%%%%%%%

As one can see from Eq.~(\ref{eqgeod}), the proper acceleration $d^{2}x^{\mu
}/ds^{2}$ is not a covariant object. In particular, its vanishing or
non-vanishing has no observer-independent meaning. In contrast, the
relative acceleration between worldlines is a covariant quantity, and its
vanishing or non-vanishing, does not depend on the frame of reference.

Consider a one-parameter congruence of curves $x^{\mu }\left( s;\lambda
\right) $, so that for each $\lambda =\lambda _{0}=$ constant, $x^{\mu
}\left( s,\lambda _{0}\right) $ satisfies Eq.~(\ref{eqgeod}). We suppose the
parametrization to be smooth, and hence we can introduce the tangent vector
fields along the trajectories of the particles as $U^{\mu }=\partial x^{\mu
}\left( s;\lambda \right) /\partial s$ and $n^{\mu }=\partial x^{\mu }\left(
s;\lambda \right) /\partial \lambda $, respectively. We also introduce the
four-vector
\begin{equation}
\eta ^{\mu }=\left[ \frac{\partial x^{\mu }\left( s;\lambda \right)
}{\partial \lambda }\right] \delta \lambda \equiv n^{\mu }\delta \lambda ,
\ee
joining points on infinitely close geodesics, corresponding to parameter
values $\lambda $ and $\lambda +\delta \lambda $, which have the same value
of $s$ \cite{LaLi, holten}. From the definition of $U^{\mu }$ and $n^{\mu }$ it follows that
they satisfy the relation $\partial U^{\mu }/\partial \lambda =\partial
n^{\mu }/\partial s$. Then it can be easily shown that $n^{\nu }\nabla _{\nu
}U^{\mu }=U^{\nu }\nabla _{\nu }n^{\mu }$ \cite{LaLi,holten}. Now consider the
second derivative
\begin{eqnarray}
\frac{D^{2}n^{\mu }}{ds^{2}} &\equiv &U^{\nu }\nabla _{\nu }\left( U^{\alpha
}\nabla _{\alpha }n^{\mu }\right) =U^{\nu }\nabla _{\nu }\left( n^{\alpha
}\nabla _{\alpha }U^{\mu }\right)
     \nonumber \\
&=&\left( \nabla _{\nu }\nabla _{\alpha }U^{\mu }\right) n^{\alpha }U^{\nu
}+\left( \nabla _{\nu }n^{\alpha }\right) \left( \nabla _{\alpha }U^{\mu
}\right) U^{\nu }. \nonumber\\ \label{eq2}
\end{eqnarray}

By changing the order of covariant differentiation by using the definition
of the Riemann curvature tensor $R_{\beta \nu \alpha }^{\mu }$, $\left(
\nabla _{\nu }\nabla _{\alpha }-\nabla _{\alpha }\nabla _{\nu }\right)
U^{\mu }=-R_{\beta \nu \alpha }^{\mu }U^{\beta }$ \cite{LaLi}, Eq.~(\ref{eq2}) can be
written as
\begin{equation}
\frac{D^{2}n^{\mu }}{ds^{2}}=R_{\nu \alpha \beta }^{\mu }n^{\alpha }U^{\beta
}U^{\nu }+\nabla _{\alpha }\left( U^{\nu }\nabla _{\nu }U^{\mu }\right)
n^{\alpha }.
\end{equation}

By taking into account Eq.~(\ref{eqgeod}), after multiplication with the
constant factor $\delta \lambda $, we obtain the geodesic deviation equation
(Jacobi equation) as \cite{LaLi,holten,Haw,Gron}
\begin{equation}
\frac{D^{2}\eta ^{\mu }}{ds^{2}}=R_{\nu \alpha \beta }^{\mu }\eta ^{\alpha
}U^{\beta }U^{\nu }+\eta ^{\alpha }\nabla _{\alpha }f^{\mu }.  \label{eq3}
\end{equation}

In the case $f^{\mu }\equiv 0$ we reobtain the standard Jacobi equation,
corresponding to the geodesic motion of test particles. The interest in the
deviation vector $\eta ^{\mu }$ derives from the fact that if $x_{0}^{\mu
}(s)=x^{\mu }\left( s;\lambda _{0}\right) $ is a solution of Eq.~(\ref{eq3}%
), then to first order $x_{1}^{\mu }(s)=x_{0}^{\mu }(s)+\eta ^{\mu }$ is a
solution as well, since $x^{\mu }\left( s;\lambda _{1}\right) \approx x^{\mu
}\left( s;\lambda _{0}\right) +n^{\mu }\left( s;\lambda _{0}\right) \delta
\lambda \approx x^{\mu }\left( s;\lambda _{0}+\delta \lambda \right) $.

Note that the geodesic deviation equation Eq.~(\ref{eq3}) can also be derived from the Lagrangian \cite{holten}
\bea
L\left( \eta \right) &=&\frac{1}{2}g_{\mu \nu }\frac{D\eta ^{\mu }}{ds}\frac{%
D\eta ^{\nu }}{ds}+\frac{1}{2}R_{\mu \nu \alpha \beta }\eta ^{\mu }\eta
^{\alpha }U^{\beta }U^{\nu }
    \nonumber\\
&&+g_{\beta \mu }\eta ^{\beta }\eta ^{\alpha
}\nabla _{\alpha }f^{\mu }.
\eea

In this Lagrangian the metric, connection and curvature are those of a given
reference geodesic $x_{0}^{\mu }(s)$, with $U^{\mu }\left( s\right) =\dot{x}%
_{0}^{\mu }\left( s\right) $ representing the four velocity along the same
geodesic. These quantities are the background variables. The $\eta ^{\mu
}\left( s\right) $ are the independent Lagrangian generalized coordinates,
which are to be varied in the action according to the Lagrange equations,
\begin{equation}
\frac{d}{ds}\frac{\partial L}{\partial \left( D\eta ^{\nu }/ds\right) }-%
\frac{\partial L}{\partial \eta ^{\nu }}=0.
\end{equation}
Then these Lagrange equations give again the equation of the geodesic
deviation.

By taking into account the explicit form of the extra-force
given by Eq.~(\ref{extra}), in modified theories of gravity with a curvature-matter coupling, the geodesic deviation equation can be written as
%\begin{widetext}
\bea
\frac{D^{2}\eta ^{\mu }}{ds^{2}}=R_{\nu \alpha \beta }^{\mu }\eta ^{\alpha
}U^{\beta }U^{\nu }
      \nonumber  \\
 +\eta ^{\alpha }\nabla _{\alpha }\left\{ \nabla _{\nu
}\ln \left[ f_{L_m}\left( R, L_m\right)\frac{%
dL_{m}\left( \rho \right) }{d\rho } \right] \left(
L_{m}g^{\mu \nu }-T^{\mu \nu}\right) \right\} \text{.} \nonumber
\eea
%\end{widetext}

Explicitly, the geodesic deviation equation becomes
\begin{widetext}
\bea
\frac{D^{2}\eta ^{\mu }}{ds^{2}}&=&R_{\nu \alpha \beta }^{\mu }\eta ^{\alpha
}U^{\beta }U^{\nu }
    +\eta ^{\alpha }\left\{\nabla _{\alpha }\nabla _{\nu }\ln %
\left[f_{L_m}\left( R,L_m\right)\frac{%
dL_{m}\left( \rho \right) }{d\rho }  \right]\right\} \left(
L_{m}g^{\mu \nu }-T^{\mu \nu }\right)
   \nonumber\\
&&+\eta ^{\alpha }\nabla _{\nu }\ln \left[ f_{L_m}\left( R,L_m\right)\frac{%
dL_{m}\left( \rho \right) }{d\rho } \right]\left( g^{\mu \nu } \nabla _{\alpha }L_{m} -\nabla
_{\alpha }T^{\mu \nu }\right) .
\eea

As a specific example, consider the case of a linear coupling between
curvature and matter \cite{Bertolami:2007gv}, where the Lagrangian given by $f(R,L_m)=f_1(R)+\lambda f_2(R) L_m$, and $f_i(R)$ (with i = 1, 2) are arbitrary functions of the Ricci scalar $R$ and the strength of the interaction between $f_2(R)$ and the matter Lagrangian is characterized by a coupling constant λ. For this case, the geodesic deviation equation can be written as
%\begin{widetext}
\begin{eqnarray}
\frac{D^{2}\eta ^{\mu }}{ds^{2}} &=&R_{\nu \alpha \beta }^{\mu }\eta
^{\alpha }U^{\beta }U^{\nu }+
\lambda \frac{\left[ 1+\lambda f_{2}(R)\right] d\left[ \ln F_{2}(R)\right]
/dR-1}{\left[ 1+\lambda f_{2}(R)\right] ^{2}}\left( L_{m}g^{\mu \nu }-T^{\mu
\nu }\right) \eta ^{\alpha }\nabla _{\alpha }R\nabla _{\nu }R
     \nonumber \\
&&+ \frac{\lambda F_{2}\left( R\right) }{1+\lambda f_{2}\left( R\right) }%
\left( L_{m}g^{\mu \nu }-T^{\mu \nu }\right) \eta ^{\alpha }\nabla _{\alpha
}\nabla _{\nu }R+
\frac{\lambda F_{2}\left( R\right) }{1+\lambda f_{2}\left( R\right) }\eta
^{\alpha }\left( g^{\mu \nu }\nabla _{\alpha }L_{m}-\nabla _{\alpha }T^{\mu
\nu }\right) \nabla _{\nu }R.
\end{eqnarray}
\end{widetext}

Note that a second order tensor $\nabla _{\nu }U_{\mu }$ can be decomposed into symmetric and antisymmetric parts, and the symmetric part can be further
decomposed into a trace and trace-free part. Thus, in general, we can write
\cite{Haw,Gron}
\begin{equation}
\nabla _{\mu }U_{\nu }=\frac{1}{3}\theta h_{\mu \nu }+\sigma _{\mu \nu
}+\omega _{\mu \nu }+\dot{U}_{\mu }U_{\nu },
\end{equation}
where $h_{\mu \nu }=g_{\mu \nu }-U_{\mu }U_{\nu }$, $\dot{U}_{\mu }=U^{\nu
}\nabla _{\nu }U_{\mu }$, $\theta =\nabla _{\nu }U^{\nu }$ is the expansion
of the congruence of particles. The shear $\sigma _{\mu \nu }$ is given by
\be
\sigma _{\mu \nu }=\nabla _{(\mu }U_{\nu )}-\frac{1}{3} \theta h_{\mu \nu }-\dot{U}%
_{(\mu }U_{\nu )},
\ee
 where
 \be
 \nabla _{(\mu }U_{\nu )}=\frac{1}{2} \left(
\nabla _{\nu }U_{\mu }+\nabla _{\mu }U_{\nu }\right),
\ee
 and the vorticity $\omega _{\mu \nu }$ is defined as
 \be
 \omega _{\mu \nu }=\nabla
_{\lbrack \mu }U_{\nu ]}-\dot{U}_{[\mu }U_{\nu ]},
\ee
 respectively, where
 \be
\nabla _{\lbrack \mu }U_{\nu ]}=\frac{1}{2} \left( \nabla _{\nu
}U_{\mu }-\nabla _{\mu }U_{\nu }\right) .
\ee
The term $\dot{U}_{\mu
}U_{\nu }$ takes into account the possible presence of other forces, which
are orthogonal to the four-velocity, with four-acceleration given by $\dot{U}%
_{\mu }=U^{\nu }\nabla _{\nu }U_{\mu }$.

From the definition of the Riemann curvature we have
\be
\left(\nabla_{\nu}\nabla _{\alpha }-\nabla _{\alpha }\nabla _{\nu }\right) U^{\mu }=-R_{\beta \nu \alpha }^{\mu }U^{\beta }  \,.
\ee
By contracting with $\mu =\nu $ and after
multiplication with $U^{\alpha }$ we obtain  \cite{Gron}
\be
 U^{\alpha }\nabla _{\nu
}\nabla _{\alpha }U^{\nu }-U^{\alpha }\nabla _{\alpha }\theta =-R_{\alpha
\beta }U^{\alpha }U^{\beta }.
\ee
The first term in this equation can be written as
\be
U^{\alpha }\nabla _{\nu }\nabla _{\alpha }U^{\nu }=\nabla _{\nu
}\left( U^{\alpha }\nabla _{\alpha }U^{\nu }\right) -\left( \nabla _{\nu
}U_{\alpha }\right) \left( \nabla ^{\alpha }U^{\nu }\right) .
\ee
Hence we obtain the Rachaudhury equation in the presence of an extra force as \cite{Haw,Gron}
\begin{equation}
\dot{\theta}+\frac{1}{3}\theta ^{2}+\left( \sigma ^{2}-\omega ^{2}\right)
=\nabla _{\mu }f^{\mu }+R_{\mu \nu }U^{\mu }U^{\nu },
\end{equation}
where $\sigma ^{2}=\sigma _{\mu \nu }\sigma ^{\mu \nu }$ and $\omega
^{2}=\omega _{\mu \nu }\omega ^{\mu \nu }$, respectively.

With the use of the field equation Eq.~(\ref{eqrmunu}), and the
expression of the extra force, in modified theories of gravity theories with an arbitrary coupling between curvature and matter, the Raychaudhury equation assumes the following generalised form
\begin{eqnarray}
\dot{\theta} &=&-\frac{1}{3}\theta ^{2}-\left( \sigma ^{2}-\omega
^{2}\right)  + \Lambda \left( R,L_{m}\right)
   \nonumber  \\
&&\hspace{-0.5cm} +\nabla _{\mu }\left\{ \nabla _{\nu }\ln \left[  f_{L_m}\left( R,L_m\right)\frac{%
dL_{m}\left( \rho \right) }{d\rho } \right] \left( L_{m}g^{\mu \nu }-T^{\mu
\nu }\right) \right\}
     \nonumber \\
&&+\frac{1}{f_R\left( R,L_{m}\right) }U^{\mu
}U^{\nu }\nabla _{\mu }\nabla _{\nu }f_R\left( R,L_{m}\right)
     \nonumber  \\
&&+\Phi \left(
R,L_{m}\right) \left( T_{\mu \nu }U^{\mu }U^{\nu }-\frac{1}{3}T\right) .
\end{eqnarray}

The vorticity $\omega _{\mu \nu }$ satisfies the equation \cite{Haw,Gron}
\begin{equation}
\dot{\omega}_{\mu \nu }=-\frac{2}{3}\theta \omega _{\mu \nu }-2\sigma
_{\lbrack \nu }^{\lambda }\omega _{\mu ]\lambda }+\nabla _{\lbrack \mu
}f_{\nu ]},
\end{equation}
while the dynamics of the shear $\sigma _{\mu \nu }$ is described by the
equation \cite{Haw,Gron}
%\begin{widetext}
\bea
\dot{\sigma}_{\mu \nu } &=& \frac{1}{2}h_{\mu }^{\sigma }h_{\nu }^{\lambda
}R_{\sigma \lambda }+h_{\mu }^{\lambda }h_{\nu }^{\sigma }\nabla _{(\lambda
}f_{\sigma )}-\frac{2}{3}\theta \sigma _{\mu \nu }-\omega _{\mu \lambda
}\omega _{\nu }^{\lambda }
    \nonumber   \\
&&-\sigma _{\mu \lambda }\sigma _{\nu }^{\lambda }-\frac{1}{3}h_{\mu \nu }\left( \omega ^{2}-\sigma ^{2}+\frac{1}{2}h^{\sigma
\lambda }R_{\sigma \lambda }+\nabla _{\mu }f^{\mu }\right)
    \nonumber  \\
&&-C_{\mu \sigma
\nu \lambda }U^{\sigma }U^{\lambda },
\eea
%\end{widetext}
where
\be
C_{\mu \sigma \nu \lambda }=R_{\mu \sigma \nu \lambda }+g_{\mu
\lbrack \lambda }R_{\nu ]\sigma }+g_{\sigma \lbrack \nu }R_{\lambda ]\mu
}+\frac{Rg_{\mu \lbrack \nu }g_{\lambda ]\sigma }}{3},
\ee
is the Weyl tensor. With the use of the field equations Eq.~(\ref{eqrmunu}) and of the expression of the extra force $f^{\mu }$, the evolution equations for the shear and vorticity can also be obtained explicitly for modified gravity with an arbitrary curvature-matter coupling.

%%%%%%%%%%%%%%%%%%%%%%%%%%%%%%%%%%%%%%%%%%%%%%%%%%%%%%%%%%%
\section{Tidal forces with an arbitrary curvature-matter coupling}\label{sect5}
%%%%%%%%%%%%%%%%%%%%%%%%%%%%%%%%%%%%%%%%%%%%%%%%%%%%%%%%%%%

Tides are the manifestation of a gradient of the gravitational force field
induced by a mass above an extended body or a system of particles. In the
Solar System tidal perturbations act on compact bodies such as planets,
moons and comets. On larger scales than the solar system, as in a galactic
or cosmological context, one can observe tidal deformations or disruptions
of a stellar cluster by a galaxy, or in galaxy encounters \cite{mas}. In the
relativistic theories of gravitation, as well as in Newtonian gravity, a
local system of coordinates can be chosen, which is inertial except for the
presence of the tidal forces. In strong gravitational fields, relativistic
tidal effects can lead to interesting phenomena, such as the emission of
tidal gravitational waves \cite{mas}. Relativistic corrections to the
Newtonian tidal accelerations caused by a massive rotating source, such as,
for example, the Earth, could be determined experimentally, at least in
principle, thus leading to the possibility of testing relativistic theories
of gravitation by measuring such effects in a laboratory.

\subsection{There are no geodesic reference frames in $f\left(R,L_m\right)$ gravity}

If the Christoffel symbols $\Gamma _{\mu \nu }^{\alpha }$ associated to a
given metric are symmetric, it is always possible to chose a coordinate
system in which all the $\Gamma _{\mu \nu }^{\alpha }$ become zero at a
previously assigned point \cite{LaLi}. The corresponding coordinates are
called Riemann normal coordinates, and the system of reference can be called
a locally Minkowski system. Let the given point be the origin of the
coordinate system, and let the initial values of the Christoffel symbols be
equal to $( \Gamma _{\alpha \beta }^{\mu })_{0}$. In the
neighborhood of the origin we introduce the coordinate transformation
\bea
x^{\prime \mu }=x^{\mu }+\frac{1}{2} \left( \Gamma _{\alpha \beta }^{\mu }\right)
_{0}x^{\alpha }x^{\beta }  \,,
\eea
which provides
\be
\left(\frac{\partial ^{2}x^{\prime
\sigma }}{\partial x^{\alpha }\partial x^{\beta }} \right)
\left(\frac{\partial x^{\prime \mu }}{\partial x^{\prime }\sigma}
\right) = \left( \Gamma_{\alpha \beta }^{\mu }\right) _{0} \,.
\ee
Then, from the transformation law of the Christoffel symbols,
\bea
\left( \Gamma _{\alpha \beta }^{\mu }\right)_{0}&=&\Gamma_{\lambda \sigma}^{\prime \nu }\left(  \frac{\partial x^{\mu
}}{\partial x^{\prime \nu }}\right)
\left( \frac{\partial x^{\prime \lambda}}{\partial x^{\alpha }}\right)
\left( \frac{\partial x^{\prime \sigma }}{\partial
x^{\beta }}\right)
  \nonumber \\
&&+\left( \frac{\partial ^{2}x^{\prime \sigma }}{\partial x^{\alpha
}\partial x^{\beta }}\right)
\left( \frac{\partial x^{\prime \mu }}{\partial
x^{\prime }\sigma} \right)
\eea
it follows that all the transformed Christoffel symbols $\Gamma _{\lambda \sigma }^{\prime \nu }$ are zero.

One can also show that by a suitable choice of the coordinate system one can make all the $\Gamma _{\alpha \beta }^{\mu }$ go to zero not only in a point, but all along a given world line \cite{LaLi}. At the same time with the vanishing of the Christoffel symbols, the metric $g_{\mu \nu }$ can be reduced, at a given point, to its diagonal (Minkowskian) form, $g_{\mu \nu }=\eta _{\mu \nu }$, and consequently in the given point the covariant derivatives coincide with the ordinary partial derivatives, $\nabla _{\mu }=\partial /\partial x^{\mu }$ \cite{LaLi,Haw,mas,Oh}.

In such a coordinate system the equation of motion Eq.~(\ref{eqgeod}) takes the form
\begin{equation}
\left.\frac{d^{2}x^{\mu}}{dt^{2}}\right|_{geod}=\left.f^{\mu }\right|_{geod},
\ee
where for small particle velocities $ds\approx dt$, where $t$ is the time coordinate.

Therefore, in theories of gravity with an arbitrary curvature-matter coupling,
despite the fact that one can locally cancel the Christoffel symbols, and can introduce a flat (Minkowskian) metric, due to the presence of the
extra-force $f^{\mu }$, which generally cannot be reduced to zero, the
motion of the test particles is non-inertial, and they will always
experience a supplementary acceleration induced by the presence of the
coupling between matter and curvature.

\subsection{Tidal forces in $f\left(R,L_m\right)$ gravity}

In the following we will denote a
reference frame in which all the Christoffel symbols vanish by a prime. In
such a system one can always take $\eta ^{\prime 0}=0$, which means that the
particle accelerations are compared at equal times. $\eta ^{\prime i}$ is
then the displacement of the particle from the origin \cite{Oh}. Moreover,
in the static/stationary case, in which the metric, the Ricci scalar and the
thermodynamic parameters of the matter do not depend on time, $\ f^{\prime
0}=0$. With the use of the equation of motion this condition implies $%
U^{\prime 0}=$ constant $=1$. Therefore, with these assumptions, the
equation of the geodesic deviation (the Jacobi equation) takes the form
\begin{equation}
\frac{d^{2}\eta ^{\prime i}}{dt^{\prime 2}}=R_{0l0}^{\prime i}\eta ^{\prime
l}+R_{jlm}^{\prime i}\eta ^{\prime l}U^{\prime j}U^{^{\prime }m}+\eta
^{\prime l}\frac{\partial f^{\prime i}}{\partial x^{\prime l}}.
\label{geodin}
\end{equation}

Equation~(\ref{geodin}) can be reformulated as
\begin{equation}
F^{\prime i}=\frac{d^{2}\eta ^{\prime i}}{dt^{\prime 2}}=K_{j}^{i}\eta
^{\prime j},
\end{equation}
where $F^{\prime i}$ is the tidal force, and we have introduced the
generalized tidal matrix $K_{l}^{i}$ \cite{mas}, which is defined as
\begin{equation}
K_{j}^{i}=R_{0j0}^{\prime i}+R_{kjm}^{\prime i}U^{\prime k}U^{^{\prime }m}+%
\frac{\partial f^{\prime i}}{\partial x^{\prime j}}.
\end{equation}
The tidal force has the property $\partial F^{\prime i}/\partial \eta
^{\prime j}=K_{j}^{i}$, and its divergence is given by $\partial F^{\prime
i}/\partial \eta ^{\prime i}=K$, where the trace $K$ of the tidal matrix is
\begin{equation}
K=K_{i}^{i}=R_{00}^{\prime }+R_{km}^{\prime }U^{\prime k}U^{^{\prime }m}+%
\frac{\partial f^{^{\prime }j}}{\partial x^{\prime j}}.
\end{equation}

With the use of the gravitational field equations Eq.~(\ref{eqrmunu}) we can
express $K$ as
\begin{widetext}
\begin{eqnarray}
K &=&\Lambda \left( R^{\prime },L_{m}^{\prime }\right) \eta _{00}+\frac{1}{%
f_R\left( R^{\prime },L_{m}\right) }\frac{\partial ^{2}}{\partial t^{\prime 2}}%
f_R\left( R^{\prime },L_{m}^{\prime }\right) +\Phi \left( R^{\prime
},L_{m}^{\prime }\right) \left( T_{00}^{\prime }-\frac{1}{3}T^{\prime }\eta
_{00}\right) + \Lambda \left( R^{\prime },L_{m}^{\prime }\right) \eta _{km}U^{\prime
k}U^{^{\prime }m}
     \nonumber \\
&&+\frac{1}{f_R\left( R^{\prime },L_{m}^{\prime }\right) }%
U^{\prime k}U^{\prime m}\frac{\partial ^{2}}{\partial x^{\prime k}\partial
x^{\prime m}}f_R\left( R^{\prime },L_{m}^{\prime }\right) +
\phi \left( R^{\prime },L_{m}^{\prime }\right) \left( T_{km}^{\prime }-%
\frac{1}{3}T\eta _{km}\right) U^{\prime k}U^{\prime m}
    \nonumber \\
&&+\frac{\partial }{\partial x^{\prime k}}\left\{ \frac{\partial }{\partial
x^{\prime m}}\ln \left[  f_{L_m}\left( R^{\prime
},L_{m}^{\prime } \right) \frac{%
dL_{m}^{\prime }\left( \rho \right) }{d\rho }\right] \left( L_{m}^{\prime }\eta ^{km}-T^{\prime km}\right)
\right\} .
\end{eqnarray}
\end{widetext}

Then, since in the Newtonian limit one can omit all time derivatives, we
obtain for $R_{0l0}^{i}$ the expression $R_{0l0}^{i}=\partial ^{2}\phi
/\partial x^{i}\partial x^{l}$ \cite{Oh}, where, for simplicity, in the
following we will omit the primes for the geometrical and physical
quantities in the Newtonian approximation. In Newtonian gravity $\tau
_{il}=-\partial ^{2}\phi /\partial x^{i}\partial x^{l}$ represents the
Newtonian tidal tensor \cite{Oh}. Therefore, in modified gravity with a
curvature-matter coupling we obtain the tidal acceleration of the test
particles as
\begin{equation}
\frac{d^{2}\eta ^{i}}{dt^{2}}=F^{i}\approx \frac{\partial ^{2}\phi }{%
\partial x^{i}\partial x^{l}}\eta ^{l}+R_{jlm}^{i}\eta ^{l}V^{j}V^{m}+\eta
^{l}\frac{\partial f^{i}}{\partial x^{l}},  \label{newgeod}
\end{equation}
where $V^{j}$ and $V^{m}$ are the Newtonian three-dimensional velocities. In
the Newtonian approximation, in modified theories of gravity with a curvature-matter coupling the tidal force tensor is defined as
\begin{equation}
\frac{\partial F^{i}}{\partial \eta ^{l}}=\frac{\partial ^{2}\phi }{\partial
x^{i}\partial x^{l}}+R_{jlm}^{i}V^{j}V^{m}+\frac{\partial f^{i}}{\partial
x^{l}},
\end{equation}
and its trace gives the generalized Poisson equation,
\begin{equation}
\frac{\partial F^{i}}{\partial \eta ^{i}}=\Delta \phi +R_{jm}V^{j}V^{m}+%
\frac{\partial f^{i}}{\partial x^{i}}.
\end{equation}

\subsection{The Roche limit in modified gravity with an arbitrary
curvature-matter coupling}

In Newtonian gravity, the spherical potential of a given particle with
mass $M$ is $\phi (r)=-M/8\pi r$. By choosing a frame of reference so that
the $x$-axis passes through the particle's position, corresponding to the
radial spherical coordinate, that is, $(x=r,y=0,z=0)$, the Newtonian tidal
tensor is diagonal, and has the only non-zero components
\be
\tau _{ii}=\mathrm{%
diag}\left( \frac{2M}{8\pi r^{3}},-\frac{M}{8\pi r^{3}},-\frac{M}{8\pi r^{3}}\right) .
\ee
The Newtonian tidal force $\vec{F}_{t}$ can be written as $F_{tx}=2M\Delta
x/8\pi r^{3}$, $F_{ty}=-GM\Delta y/8\pi r^{3}$ and $F_{tz}=-M\Delta z/8\pi
r^{3}$, respectively \cite{Oh}. These results can be used to derive the
generalization of the Roche limit in modified gravity with an arbitrary coupling between matter and curvature.

The Roche limit is the closest distance $r_{Roche}$ that a celestial object
with mass $m$, radius $R_{m}$ and density $\rho _{m}$, held together only by
its own gravity, can come to a massive body of mass $M$, radius $R_{M}$ and
density $\rho _{M}$, respectively, without being pulled apart by the massive
object's tidal (gravitational) force \cite{mas}. For simplicity we will
consider $M \gg m$, so that the center of mass of the system nearly coincides
with the geometrical center of the mass $M$.

The elementary Newtonian theory of this process is as follows. Consider a small mass $\Delta m$ located at the surface of the small object of mass $m$. There are two forces acting on $\Delta m$, the gravitational attraction of the mass $m$, given by
\be
F_{G}=\frac{m\Delta m}{8\pi R_{m}^{2}},
\ee
and the tidal force exerted by the massive object, which is given by
\be
F_{t}=\frac{M\Delta mR}{8\pi r^{3}},
\ee
where $r$ is the distance between the centers of the two celestial bodies. The Roche limit is reached at the distance $r=r_{Roche}$, when the gravitational force and the tidal force exactly balance each other, $F_{G}=F_{t}$, thus giving \cite{mas}
\be
r_{Roche}=R_{m}\left(\frac{ M}{m}\right) ^{1/3}=2^{1/3}R_{M}\left( \frac{\rho _{M}}{\rho
_{m}}\right) ^{1/3} .
\ee

In modified gravity with a curvature-matter coupling the equilibrium
between gravitational and tidal forces occurs at a distance $r_{Roche}$ given by the equation
\begin{equation}
\left( \frac{M}{8\pi r_{Roche}^{3}}+R_{jrm}^{r}V^{j}V^{m}+\frac{\partial
f^{r}}{\partial r}\right) R_{m}=\frac{m}{8\pi R_{m}^{2}}+f^{r},
\end{equation}
where $f^{r}$ is the radial component of the extra-force, which modifies the
Newtonian gravitational force, and the curvature tensor $R_{jrm}^{r}$ (no
summation upon $r$) must be evaluated in the coordinate system in which the
Newtonian tidal tensor is diagonal. Hence we obtain the generalized Roche
limit in the presence of arbitrary geometry-matter coupling as
%\begin{widetext}
\begin{eqnarray}
r_{Roche}&\approx & R_{m}\left( \frac{M}{m}\right) ^{1/3} \times
    \nonumber \\
&&\hspace{-1.75cm}\times \left[ 1+\frac{8\pi
R_{m}^{3}}{3m}\left( R_{jrm}^{r}V^{j}V^{m}+\frac{\partial f^{r}}{\partial r}%
\right) -\frac{8\pi R_{m}^{2}}{3m}f^{r}\right] ,\nonumber\\
\end{eqnarray}
%\end{widetext}
where we have assumed that the gravitational effects due to the coupling
between matter and curvature are small as compared to the Newtonian ones.

%%%%%%%%%%%%%%%%%%%%%%%%%%%%%%%%%%%%%%
\section{Conclusions}
%%%%%%%%%%%%%%%%%%%%%%%%%%%%%%%%%%%%%%

In the present paper we have developed some of the basic theoretical tools
necessary to investigate the properties of gravitational models with direct
curvature-matter interaction. In particular, we have obtained the
Raychaudhury equation, which can be used to analyze the existence of black
holes in these types of theories, and to rigorously formulate the
singularity theorems. The presence of the extra-force in the gravitational model leads to the appearance of some new terms in the Raychaudhury and the geodesic deviation equation, terms which are a direct consequence of the curvature-matter coupling. Therefore the presence of the extra force can have a significant effect on the formation of massive astrophysical objects by gravitational collapse.  However, we have to mention in the present gravity model the full
non-linear theory contains higher than two spacetime derivatives in the field equations. Hence  modified gravity theories with geometry-matter coupling can be considered as effective theories only apart from the special Einstein case.

The geodesic equation can be used to study the effects of the generalized tidal forces, which could lead to the possibility of observationally testing the model through the observational effects of the tides due to an extended mass distribution. Typical situations in which the effects of the tides are of major importance are galactic encounters, globular clusters under the influence of the galactic mass distribution, and Oort cloud perturbations by the galactic field \cite{tide}. It was suggested that the compressive tidal field in the center of the flat core of early type galaxies and of the ultra-luminous galaxies compresses the molecular clouds, producing the dense gas observed in the center of these galaxies. The curvature-matter interaction modifies the nature of the tidal forces. Therefore the observational study of the tidal forces may give some insights in the fundamental aspects of the gravitational interaction.

On the other hand the Newtonian limit of the geodesic deviation equation
Eq.~(\ref{newgeod}) shows that the tidal motion of test particles is
directly influenced not only by the gradient of the extra force $f$, which
is basically determined by the gradient of the Ricci scalar, but also by an
explicit coupling between the velocity and the Riemann curvature tensor. The
possibility that the Pioneer anomaly \cite{pion} is due to the presence of
the extra force due to the geometry-matter coupling was analyzed in \cite
{Bertolami:2007gv}. However, Eq.~(\ref{newgeod}%
) shows that this effect may also be velocity dependent, leading to the
possibility of the detection of these type of extra gravitational effects by
the method of the Doppler tracking of a spacecraft. On the other hand, recent researches on the Pioneer anomaly \cite{therm}  have given convincing
arguments against its modified gravity origin. 

To conclude, the present study  opens further
possibilities for the experimental and observational investigation of the possible coupling
between curvature and matter, and for a large class of modified gravity theories. 

\section*{Acknowledgments}

FSNL acknowledges financial support of the Funda\c{c}\~{a}o para a Ci\^{e}ncia e Tecnologia through the grants CERN/FP/123615/2011 and CERN/FP/123618/2011.

\end{document}